\documentclass[twocolumn,showpacs]{revtex4}
\usepackage{graphics,epsfig,graphicx}
\usepackage{color}
\usepackage{latexsym}
\usepackage{calrsfs}
\begin{document}


\title{Quantum Signature Blurred by Disorder in Indirect Exciton
Gases}

\author{Mathieu Alloing$^{1}$, Aristide Lema\^{i}tre$^{2}$, Fran\c{c}ois
Dubin$^{1}$}

\affiliation{$^{1}$ ICFO-Institut de Ci\`{e}ncies Fot\`{o}niques,
Mediterranean Technology Park, Av. del Canal Ol\'{i}mpic, E-08860
Castelldefels, Spain}
\affiliation{
$^{2}$ Laboratoire de Photonique et Nanostructures, LPN/CNRS, Route de
Nozay, 91460 Marcoussis, France}

\date{\today }
\pacs{78.67.De, 73.63.Hs}

\begin{abstract}
The photoluminescence dynamics of a microscopic
gas of indirect excitons trapped in coupled quantum wells is probed at very low
bath temperature ($\approx$ 350 mK). Our experiments reveal the non linear
energy relaxation characteristics of indirect excitons. Particularly, we observe
that the excitons dynamics is strongly correlated with the screening of
structural disorder by repulsive exciton-exciton interactions. For our
experiments where two-dimensional excitonic states are gradually defined, the
distinctive enhancement of the exciton scattering rate towards lowest energy
states with increasing density does not reveal unambiguously quantum statistical
effects such as Bose stimulation.

\end{abstract}

\maketitle

The development of trapping and
cooling techniques of atomic gases has lead to major advances in the
studies of quantum matter. For bosonic atoms, a striking example is given by
the Bose stimulation towards occupation of the system ground-state. This
manifests that quantum statistics is dominant and triggers the subsequent
Bose-Einstein condensation \cite{Miesner_98}. At present, atomic
condensates are routinely produced to investigate further many-body quantum
states \cite{Bloch_08}, and signatures observed with these
shall constitute milestones to demonstrate Bose-Einstein condensation of any
bosonic system.

In the solid-state, a variety of quasi-particles with bosonic character can
also be found. Particularly, semiconductors exhibit bound electron-hole
pairs called excitons, as well as (exciton-)polaritons which correspond to the
coherent superposition of photonic and excitonic degrees of freedom. For the
latter composite states, Bose amplification induced by stimulated energy
relaxation was reported \cite{Savvidis_00,Huang_00,Saba_01}, followed by the
observation of polaritons condensation \cite{Deng_02,Kasprak_06,Balili_07}.
On the other hand, signatures of Bose-Einstein statistics for
excitons are controversial \cite{Moskalenko_00}, however, quantum signatures
such as macroscopic coherence have been reported \cite{Yang_06,Timofeev_08}.

In the quest for exciton condensation, bilayer semiconductor heterostructures
were
introduced \cite{Lozovik_76} and
have resulted in significant achievements towards observation of quantum
statistics of exciton many-body states. A notable example is given by devices
where a double quantum well (DQW) is
embedded in a Schottky junction \cite{Butov_04,Butov_07}. The latter is used to
control the electric
field in the structure such that excitons are rendered indirect, i.e.
that electrons and holes are each
confined in a distinct quantum well. This situation yields important
advantages: Mostly,
the spatial overlap between the electron and hole wave-functions is greatly
reduced which leads to long lived excitonic states, from tens of nanoseconds to
microseconds timescales. On the other hand excitons can relax with a
(sub-)nanosecond
dynamics close to the lattice temperature.
In addition, excitons are electrically
polarized with well aligned electric dipoles which induces strong
exciton-exciton repulsions limiting ionization at low temperature.

Recently, DQW structures have been utilized to implement electrostatic traps for
indirect excitons. These are based on micro-patterned gate electrodes
\cite{Hammack_06,Chen_06,Gartner_06} which allow
one to confine an exciton gas in a microscopic region. In the past years, the
exciton trap technology has rapidly evolved
\cite{Remeika_09,High_09,Vogele_09} and a set of
phenomena expected for exciton condensates have been observed,
e.g. the appearance of macroscopic coherence across an exciton gas along with a
spectral narrowing and linear polarization of the fluorescence emission
\cite{Timofeev_08}. Nevertheless, direct evidences for the quantum statistics of
excitons have not been reported in exciton traps yet.

The photoluminescence dynamics of indirect excitons holds essential informations
to identify signatures of many-body quantum states. This was well illustrated in
studies of macroscopic gases which have revealed a sharp enhancement (or
``jump'') of the photoluminescence signal once the external laser excitation
creating excitons is switched off \cite{Butov_99}. In fact, in a regime free
from external heating source, indirect excitons experience a fast relaxation
towards lowest energy states which leads to a high occupation of the excitonic
ground state. Furthermore, the photoluminescence enhancement occurs at a rate
which increases with the exciton concentration such that it was interpreted as a
signature of stimulated exciton scattering \cite{Butov_01}. However, theoretical
analysis following these observations stressed the role that disorder can play
in the increase of the photoluminescence enhancement as a function of the
density  \cite{Ivanov_04}.

In this letter, we study experimentally the mutual influence of disorder and
repulsive exciton-exciton interactions in the photoluminescence dynamics of
indirect excitons. We probe a microscopic trap in which indirect excitons are
created by non resonant pulsed excitation. Besides revealing the
photoluminescence enhancement characteristics of indirect excitons, our
experiments point out strong correlations between the excitons dynamics and the
screening of disorder by repulsive exciton-exciton interactions.
As the exciton concentration is increased, repulsive dipolar interactions induce
a diffusion of excitons which become delocalized as marked by a spectral
narrowing of the photoluminescence emission. Accordingly, a clear boundary is
gradually induced between optically active and inactive exciton states that
governs the magnitude of photoluminescence enhancement \cite{Ivanov_04}. In that
context, we conclude that the excitons quantum statistics is not unambiguously signaled.

\begin{figure}
\includegraphics[width=8.5cm]{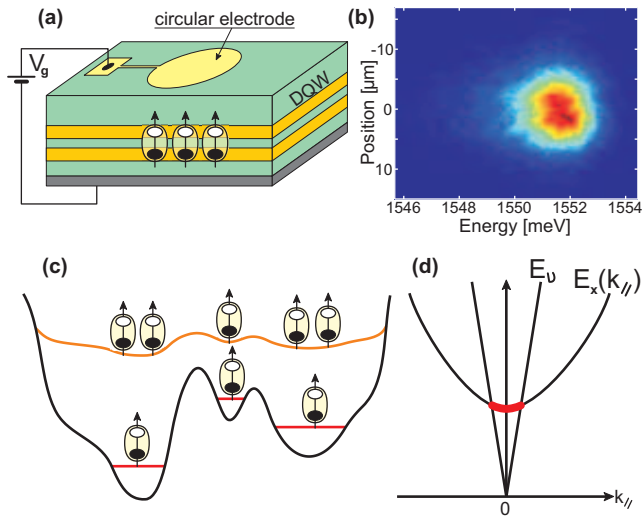}
\caption{(Color Online) (a): Schematic representation of the electrostatic trap.
Electrons and holes are displayed by filled and open circles respectively while
the arrows indicate the excitons electric dipole. (b): Spatially resolved
emission spectrum of indirect excitons recordered in a 10 ns window at low
density ($n_{\mathrm{2D}}\approx10^9$ cm$^{-2}$) and T$_\mathrm{b}$= 360 mK.
(c): Cartoon depiction of disorder screening induced by repulsive dipole
interactions: At low densities, excitons occupy minima of the random potential
while at higher densities the excitonic diffusion leads to a smooth in-plane
effective potential, i.e. that inhomogenous broadening is largely suppressed.
(d) Excitons and photons energy dispersion, E$_\mathrm{X}$ and E$_\mathrm{\nu}$
respectively. Bright exciton states, i.e. those yielding photo-emission, are
marked in bold.}
\end{figure}

As shown schematically in Fig. 1.a, we study an electrostatic trap
for indirect excitons which is generated by a semitransparent disk-shape gate
electrode of 100 $\mu$m diameter. A DQW consisting of two 8 nm wide GaAs quantum
wells separated by a 4 nm Al$_{0.33}$Ga$_{0.67}$As barrier is positioned 950
nm under the gate electrode and 50 nm above a bottom GaAs substrate rendered
homogeneously conductive by Si-doping ($n_{Si}\approx 10^{18}$ cm$^{-3}$).
This design hence fulfills requirements for a stable trapping of a
high-density exciton gas \cite{Rapaport_06}. The semiconductor sample is placed
on the He$^{3}$ insert of a He$^{4}$ optical cryostat which allows us to
perform microscopy at low bath temperature ($T_b\geq$ 350 mK).  In the
following
experiments, a constant bias V$_\mathrm{g}$= 1.33 V is applied to the gate
electrode such that the excitonic ground state is of indirect type with
holes
confined in the top quantum well and electrons in the lower one (see Fig. 1.a).
The electrostatic trap is imaged utilizing an aspherical lens, of numerical
aperture NA=0.6, which is positioned by piezo-electric transducers inside
the cryostat in a configuration where we achieve an overall spatial resolution
of
$\approx$ 1-2 $\mu$m. ``Hot`` indirect excitons are created at the center of
the trap after excitation by a defocussed laser  at 642 nm, i.e. $\approx$ 0.4
eV above the indirect excitons energy. At the trap surface, the laser excitation
exhibits a gaussian spatial
profile with 10 $\mu$m FWHM. In the
following experiments, we utilize  50 ns long laser pulses at a repetition rate
of 4 MHz and analyze
the variation of the photoluminescence (PL) of indirect excitons as a function
of the density. The latter is controlled by the power of laser excitation 
which we vary from $\approx$ 100 nW to 100 $\mu$W.  The excitonic fluorescence
is finally recorded with an
imaging spectrometer coupled to an intensified CCD camera combining 2 ns and 100
$\mu$eV temporal and spectral resolution respectively. Figure 1.b displays a
spatially resolved emission spectrum recorded at very low excitation power
($\approx$ 100 nW).

As the excitation power increases, the maximum of the photoluminescence shifts
towards higher energies. This variation results from the
repulsive dipole-dipole interactions between indirect excitons and the energy
shift, $\delta E$, gives an estimation of the gas density inside the trap
\cite{Tabou_01}. Indeed, the mean field energy associated to repulsive
exciton-exciton interactions may be expressed as
$U_\mathrm{rep}$=$u_0.n_\mathrm{2D}$, where $u_0$ is
a constant factor controlled by the DQW geometry and the correlations
between excitons \cite{Schindler_08,Rapaport_09}. For our trap, we estimate that
6 10$^{9}$ cm$^{-2}$ $\leq n_\mathrm{2D}\leq$ 8 10$^{10}$ cm$^{-2}$
for $\delta E=$ 1 meV which is obtained at an excitation power of
$\approx$10$\mu$W.

In the regime of low excitonic density,
i.e. for $n_\mathrm{2D}\approx$ 10$^{9}$ cm$^{-2}$, we note
that indirect excitons remain confined within
the region of the trap which is laser excited (see Fig. 2.b). At the same time,
the spectral distribution of the emission is rather
broad and exhibits a full width at half
maximum $\Gamma$(0)=2.70(5) meV (see Fig. 2.a). These observations
both indicate that indirect excitons are trapped by the
random fraction of the trapping potential,
U$_\mathrm{rand}(\mathrm{r}_\parallel)$, $\mathrm{r}_\parallel$ denoting the
coordinates in the plane of the quantum wells. At the center of the
trap, the electrostatic potential generated by the gate electrode is {\it a
priori} constant such that U$_\mathrm{rand}$ may be governed solely by
hetero-interface roughness and in plane
fluctuations of the electric field, for instance induced by charged impurities.
In the following, we show that exciton interactions allow us to circumvent the
limitations imposed by U$_\mathrm{rand}$.

\begin{figure}
\includegraphics[width=8.5cm]{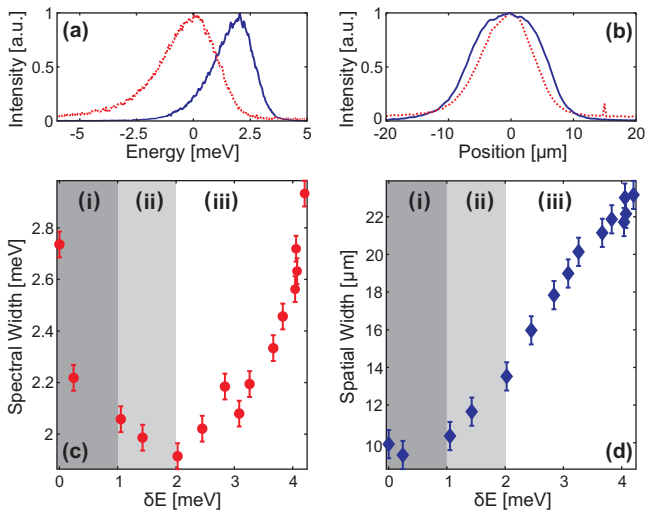}
\caption{(Color Online) (a): Emission spectrum averaged in a 20 $\mu$m wide
region at the center of the trap for $\delta E$=0 and 2 meV, dotted and solid
lines respectively. The zero in energy is taken at the center of the emission
for the lowest exciton density, i.e. for $\delta E$=0. (b): Spatial profiles of
the emissions shown in (a). (c)-(d): Spectral and spatial emission linewidths as
a function of the exciton-exciton repulsive interaction energy $\delta E$.
Experimental results were recorded in a 10 ns time window and in the same measurements performed at
T$_\mathrm{b}$= 360 mK. Error bars display the instrumental resolution.}
\end{figure}

In Figures 2.c-d, we study the variations of the emission spectral width
$\Gamma$ together with its
spatial extension $L$ as a function of $\delta E$, i.e. as a function of the
exciton concentration. The following regimes can be identified: for $\delta E
\leq$~1~meV (region (i)), $\Gamma$ decreases abruptly from 2.70(5) meV down to
2.08(5) meV while the spatial extension of the gas remains equal to
$\approx$ 10~$\mu$m. The spectral narrowing observed in this regime is a
consequence of the screening of local disorder by repulsive dipole-dipole
interactions. This behavior was predicted theoretically
\cite{Ivanov_02,Zimmerman_97}: the exciton confining potential is dressed by
$U_\mathrm{rep}$ which yields an effective in-plane potential
U$_\mathrm{eff}(\mathrm{r}_\parallel)\sim U_\mathrm{rep}(\mathrm{r}_\parallel)
+\mathrm{U}_\mathrm{rand} (\mathrm{r}_\parallel)$. Thereby, the accumulation of
indirect excitons at minima of
 U$_\mathrm{rand}$ and the depletion around its maxima smooths the spatial
profile of U$_\mathrm{eff}$ (see Fig.1.c) and the inhomogeneous broadening is
locally suppressed. After this regime,
i.e. for 1meV $\leq\delta E\leq$ 2meV (region (ii)), we observe that the
emission linewidth continues to decrease from 2.10(5)
to 1.90(5) meV while the spatial extension of the gas steadily increases from
approximately 10 to 14 $\mu$m. Hence, the exciton concentration has passed the
critical density for which repulsive dipole-dipole interactions are sufficient
to drive a macroscopic diffusion. The latter then allows for a screening of
disorder correlated at
a length scale longer than in region (i). Thereby, the inhomogenous broadening
is further reduced and the spatial extension of the cloud rapidly grows since
excitons occupy effectively delocalized states in region (ii). Let us note that
the variation of the spatial extension of the exciton cloud in region (i)-(ii)
is very similar to what observed recently in an electrostatic lattice
\cite{Remeika_09}.

In the last region of Fig. 2.c-d, i.e. in region (iii) where 2meV $\leq\delta
E\leq$ 4.2 meV, $L$ increases steadily up to 23~$\mu$m for $\delta E\approx$~4~meV,
while $\Gamma$ now increases up to $\sim$3~meV. As
shown in previous studies of similar devices \cite{High_09_NL,Zoros_09}, the
variation of $\Gamma$ signals a monotonous increase of the homogeneous
broadening as the exciton density is increased, in a regime where excitons
occupy delocalized states. At the same time, the dipolar pressure is strong for
the density range of region (iii) and the exciton cloud hence rapidly expands
\cite{Voros_06}. In addition, let us note that the highest 
exciton density which we can estimate \cite{Schindler_08,Rapaport_09} remains about an order of magnitude smaller
than the critical phase-space filling ($1/a^2$), $a$ being the exciton Bohr radius. Therefore, exciton dissociation shall not be significant in our experiments, even at large blue shifts as in region (iii)  \cite{Snoke_08}.  Finally from the variations of $\Gamma$ and $L$
presented in
Fig. 2.c-d we conclude that a sufficient level of inhomogeneous broadening is
required to observe a line narrowing, i.e. a screening of disorder by repulsive
exciton interactions (regions (i) and (ii)). In fact, previous studies performed
on heterostructures with very low inhomogeneous broadening
\cite{High_09_NL,Zoros_09} only revealed variations for $\Gamma$ which are
qualitatively what presented in region (iii) of Fig.2.c.

In the following, we analyze the dynamics of
photoluminescence and study the thermalization of indirect excitons: For our
experiments performed under non resonant laser excitation,
"hot'' indirect excitons are injected in the trap, i.e. excitons with an energy
$\approx$ 20 meV above the bottom of the energy band. On
the other hand, the detected photoluminescence
is exclusively due to the recombination of ''cold''
indirect excitons which have relaxed at the bottom of the energy band, in the
optically active region which is restricted to a maximum kinetic energy of
$\approx$ 0.1 meV (see Fig. 1.d). 

\begin{figure}
\includegraphics[width=8.5cm]{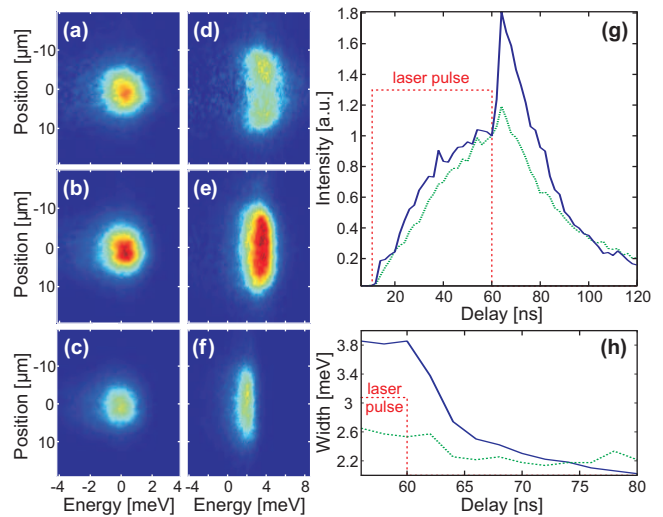}
\caption{(Color Online) (a)-(c): Spatially resolved emission spectrum measured
at $T_\mathrm{b}=$~400~mK and for $\delta E=$~0.25~meV, at the falling edge of
the laser pulse (a), and 4~ns and 20~ns later, (b) and (c) respectively. The
images were acquired with a 2~ns time window.  The energy scale corresponds to
the normalized detection energy $E-E(\delta E=0)$, where $E(\delta E=0)$ is the
energy position of the PL maximum at very low excitation power. (d)-(f): Images
recorded in the same experiment as for (a)-(c) but for $\delta E=$3.8 meV. (g):
Time evolution of the maximum of indirect excitons photoluminescence for $\delta
E=$~0.25 and 3.8 meV, dotted and solid lines respectively. (h): Time dependence
of the spectral linewidth for $\delta E=$ 0.25 and 3.8 meV, dotted and solid
lines respectively.}
\end{figure}

In Figure 3, the spatially and time resolved photoluminescence signal is
presented for low and high exciton densities,
namely for $\delta E$=0.25 and $\delta E=$~3.8 meV respectively. Depending on
the density, the system exhibits two very different behaviors after the laser
excitation is switched off. At low density ($\delta E\sim$~0.25~meV), the
linewidth decreases slightly (Fig. 3.h) while the luminescence signal drops
rapidly after a small rise (Fig. 3.g). The response to the laser interruption in
the high density regime ($\delta E=$~3.8 meV) is far more striking. The
linewidth decreases by almost a factor 2 after $\sim$5~ns while the luminescence
signal exhibits a pronounced enhancement (or "jump") by a factor $\sim$1.8 (Fig.
3.g) within the first nanoseconds. The PL signal then decreases over a few tens
of nanoseconds along with a shift towards lower energies (Fig. 3.f). This
behavior is very similar to the one observed by Butov \textit{et al.}
\cite{Butov_99} on macroscopic gases. In general, the drop of the emission
linewidth indicates a strong reduction of the effective temperature of the
indirect exciton gas once "hot" carriers are not brought anymore into the
system. It also reveals a fast thermalization of indirect
excitons in the semiconductor matrix, within a few nanosecond timescale.
Furthermore, the large enhancement of the photoluminescence marks an important
increase of the exciton population in the optically active states. Indeed,
theoretical calculations indicate for the experiments displayed in Fig. 3.d-f
that the
exciton ground state is degenerate \cite{Ivanov_04}. On the other hand, in the
dilute regime, though the effective temperature of indirect excitons is
reduced, the occupation of bright exciton states is not sufficiently enhanced
to induce a significant jump of the PL signal. In
fact, the magnitude of the photoluminescence enhancement depends strongly
on the exciton density which may be related to Bose stimulation \cite{Butov_01}.

\begin{figure}
\includegraphics[width=8.5cm]{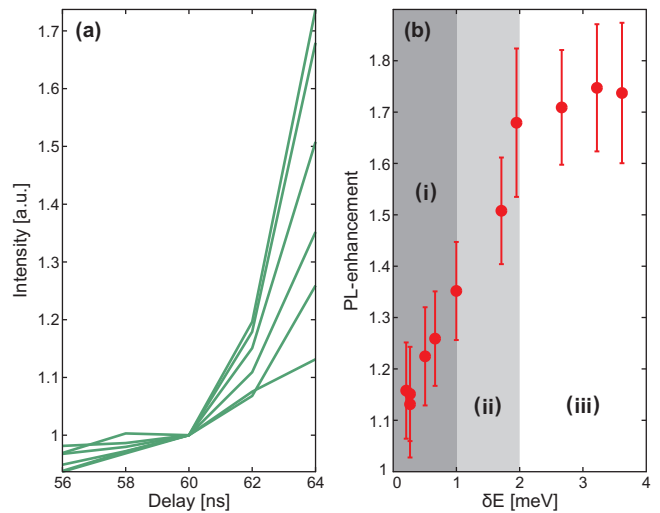}
\caption{(Color Online) (a): Time evolution of the photoluminescence enhancement
as a function of $\delta E$. From bottom to top the curves correspond to $\delta
E$=0.25-0.65-1-1.7-2-3.6 meV. (b): Magnitude of the photoluminescence
enhancement as a function of $\delta E$. These results are taken from the same
experiment as for Fig. 3 and the regions (i)-(iii) are defined as in Figure 2. Error bars display the Poissonian uncertainty for our measurements.}
\end{figure}

To identify the relation between quantum statistics and the enhancement of
photoluminescence discussed above, we studied in more details the amplitude and
the dynamics of the PL jump as a function of $\delta E$,
i.e. as a function of the exciton
density. Experimental results are displayed in Figure 4 and exhibit strong
correlations with the variations 
presented in Figure 2. Indeed, in the high density regime where the
heterostructure disorder is efficiently suppressed (region (iii) where $\delta
E\geq$ 2 meV), the rise time and the magnitude of photoluminescence
enhancement barely depend on the exciton density, unlike for
Bose stimulation.
On the other hand, for lower exciton concentrations the dynamics is far more
striking. Precisely, when the mean field energy $U_\mathrm{rep}$ is not
sufficiently large to screen the semiconductor disorder, i.e. in region (i)
where $\delta E\leq$ 1meV,  we
note a steady increase (decrease) of the amplitude (rise time) as a function
of the exciton density. Such variations contrast with theoretical works which
predict that amplitude and rise time hardly vary with the exciton density in
this regime \cite{Ivanov_04}. However, in region (i) excitons are
gradually
delocalized with
increasing $U_\mathrm{rep}$ such that a well
defined energy minimum and density of states are progressively established.
Hence, a clear distinction between optically
active and inactive exciton states is gradually induced. In this context, the
PL enhancement is governed by the interplay between the amplitude of disorder
and the strength of repulsive
exciton interactions which controls the fraction of excitonic states
contributing to
the photoluminescence emission. On the other hand, the characteristics of the PL
jump are more
intriguing for intermediate exciton concentrations. Particularly, in region (ii)
we note that the magnitude (rise time) of the photoluminescence enhancement
increases (decreases) with the exciton density. As shown in Figure 2, excitons
occupy delocalized states in this regime such that the influence of disorder can
not
account for the discrepancy between experimental results and theoretical
expectations. Therefore, the increased scattering rate towards low energy states
with increasing exciton concentration might reveal bosonic stimulation of
exciton scattering in region (ii), as discussed in Ref. \cite{Butov_01}.
However, the results displayed in Figure 4 show that amplitude
and rise time of the PL enhancement
exhibit variations in region (i) and (ii) which do not allow us to clearly
distinguish regimes where excitons are mostly
localized and delocalized respectively. Thereby, the quantum statistics of
excitons is not revealed unambiguously.

To summarize, we have studied the photoluminescence dynamics of a microscopic
gas of indirect excitons at very low bath temperature. It was shown that
repulsive exciton-exciton interactions provide an efficient screening of
structural disorder which yields a significant narrowing of the emission
spectrum accompanied by a delocalization of indirect excitons.
The photoluminescence dynamics also reveals the interplay between disorder and
repulsive dipole interactions between indirect excitons. Particularly,
increasing the exciton density, a sharp increase of the photoluminescence signal
builds up gradually in a fashion controlled by the efficiency at which disorder
is screened in the heterostructure.
Though our experiments are carried out under conditions where the exciton ground
state is highly occupied, we conclude that direct
signatures of Bose statistics are not unambiguously resolved by the
photoluminescence enhancement.

The authors are grateful to Y. Lozovik and to L. Butov for a critical reading of the manuscript,
and to M. Cristiani for stimulating discussions. F.D. and M.A. would also like to
thank J. Eschner and M. Lewenstein for their support of this project.
This work was supported partially by the the Spanish MEC (QOIT, CSD2006-00019; QNLP, FIS2007-66944) while F.D. also
acknowledges the Ram\'{o}n y Cajal program.

\end{document}